\documentclass[]{aa}
\usepackage{txfonts}
\usepackage{dblfloatfix}
\usepackage{natbib}
\usepackage{xcolor}
\usepackage{graphicx}

\newcommand{\jgrs}{   {\it J. Geophys. Res. (Space Phys.)}}

\begin{document}

\title{Recurrent Coronal Jets Induced by Repetitively Accumulated Electric Currents}

\author{Y. Guo\inst{1,2} \and P. D\'emoulin\inst{3} \and B. Schmieder\inst{3} \and M. D. Ding\inst{1,2} \and S. Vargas Dom\'inguez\inst{4} \and Y. Liu\inst{5}}

\institute{School of Astronomy and Space Science, Nanjing University, Nanjing 210093, China \\ \email{guoyang@nju.edu.cn}
\and Key Laboratory of Modern Astronomy and Astrophysics (Nanjing University), Ministry of Education, Nanjing 210093, China
\and LESIA, Observatoire de Paris, CNRS, UPMC, Universit\'e Paris Diderot, 5 place Jules Janssen, 92190 Meudon, France
\and Departamento de F\'isica, Universidad de Los Andes, A.A. 4976, Bogot\'a, Colombia
\and W. W. Hansen Experimental Physics Laboratory, Stanford University, Stanford, CA 94305, USA}


\begin{abstract}
{
Jets of plasma are frequently observed in the solar corona. A self-similar recurrent behavior
is observed in a fraction of them.
}
{
Jets are thought to be a consequence of magnetic reconnection, however, the physics involved
is not fully understood. Therefore, we study some jet observations with unprecedented temporal
and spatial resolutions.
}
{
The extreme-ultraviolet (EUV) jets were observed by the Atmospheric Imaging Assembly (AIA) on
board the \textit{Solar Dynamics Observatory} (\textit{SDO}). The Helioseismic and Magnetic
Imager (HMI) on board \textit{SDO} measured the vector magnetic field, from which we derive
the magnetic flux evolution, the photospheric velocity field, and the
vertical electric current evolution. The magnetic configuration before the jets is derived by
the nonlinear force-free field (NLFFF) extrapolation.
}
{
Three EUV jets recurred in about one hour on 2010 September 17 in the following magnetic polarity
of active region 11106. We derive that the jets are above a pair of parasitic magnetic bipoles which are continuously driven
by photospheric diverging flows. The interaction drove the build up of electric currents that we
indeed observed as elongated patterns at the photospheric level. For the first time, the high
temporal cadence of HMI allows to follow the evolution of such small currents. In the jet region,
we found that the integrated absolute current peaks repetitively in phase with the 171~\AA\ flux
evolution. The current build up and its decay are both fast, about 10 minutes each, and the current
maximum precedes the 171~\AA\ by also about 10 minutes. Then, HMI temporal cadence is marginally fast
enough to detect such changes.
}
{The photospheric current pattern of the jets is found associated to the quasi-separatrix layers
deduced from the magnetic extrapolation. From previous theoretical results, the observed diverging
flows are expected to build continuously such currents. We conclude that magnetic reconnection occurs
periodically, in the current layer created between the emerging bipoles and the large scale active region field.
It induced the observed recurrent coronal jets and the decrease of the vertical electric current magnitude.
}

\end{abstract}


\titlerunning{Recurrent Jets Induced by Accumulated Electric Currents}
\authorrunning{Guo, Y. et al.}
\maketitle

\section{Introduction} \label{sec:intro}

Solar jets are observed at many wavelengths, such as H$\alpha$, UV, EUV, and X-rays \citep{
1988Schmieder,1992Shibata173,1996Canfield,1999Chae,2012Uddin}. They are called surges in cold spectral lines (e.g., H$\alpha$)
and jets at other wavelengths, which indicates that both cold \citep[e.g.,][]{2011Srivastava,2013Kayshap} and hot plasma are accelerated
in the jet phenomenon. Magnetic reconnection is widely accepted as the central physical mechanism.
There are different magnetic reconnection models for coronal jets according to photospheric
flow patterns and coronal magnetic topologies, for instance, the emerging flux model
\citep{1977Heyvaerts,1992Shibata265,2009Gontikakis,2010Archontis}, the converging flux model \citep{1994Priest},
the Quasi-Separatrix Layer (QSL) model \citep{1996Mandrini}, and the null-point and fan-separatrix
model \citep{2008Moreno-Insertis,2009Torok,2009Pariat,2010Pariat}.
\citet{2008Moreno-Insertis} and \citet{2009Torok} use emerging flux, while \citet{2009Pariat,2010Pariat}
use horizontal photospheric twisting motion  to drive the magnetic reconnection in their simulations.
Although some of the aforementioned models focus on solar flares and coronal bright points, the associated jet
phenomenon has also been discussed.

\begin{figure*}
\centering
\includegraphics[width=1.0\textwidth]{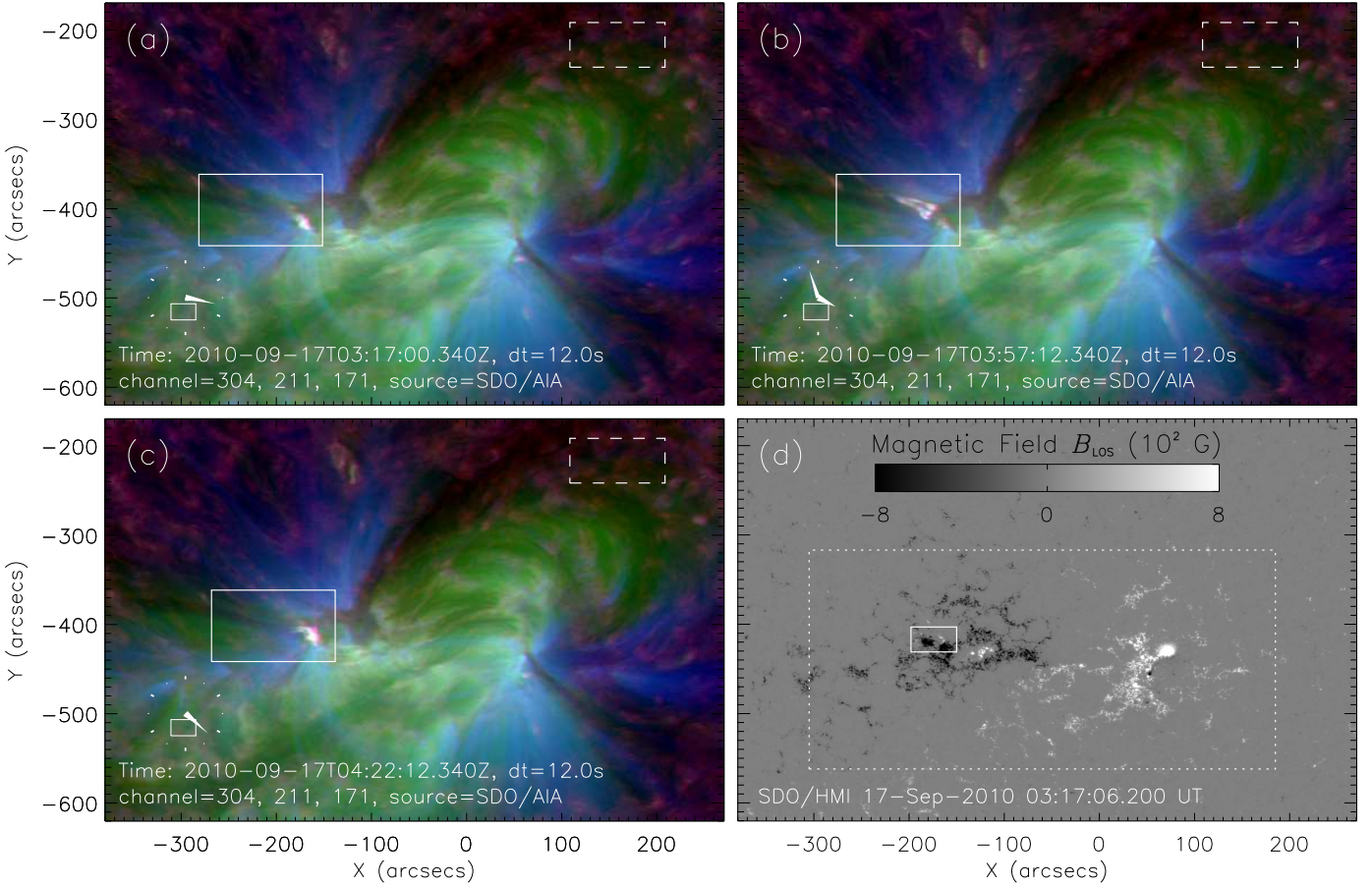}
\caption{\textbf{(a)--(c)} \textit{SDO}/AIA EUV composite images. Each EUV image consists of three
channels (red, blue, and green) in the 304~\AA , 171~\AA , and 211~\AA\ channels, whose characteristic
temperature corresponds to 0.05 MK, 0.6 MK, and 2 MK, respectively. The larger solid box marks the jet region
where the 171~\AA\ fluxes are integrated. The smaller solid box on the clock indicates the position and
field of view of the EUV image on the solar disk. The dashed box indicates the region where the background
flux for the 171~\AA \ band is computed. \textbf{(d)} \textit{SDO}/HMI line-of-sight magnetic
field. The black/white color represents the negative/positive polarity. The solid box marks the
region of the foot-points of the jets, where the line-of-sight magnetic field flux is integrated. The
dotted box marks the field of view for the analysis of the vector magnetic field.
} \label{fig:euv}
\end{figure*}

H$\alpha$ surges and EUV/X-ray jets emanating from the same region tend to appear recurrently \citep{
1995Schmieder,2001Asai,2008Chifor,2012Wang,2012Zhang}. The period ranges from tens of minutes to several hours.
It is still not clear which physical mechanism accounts for the quasi-periodicity.
\citet{2010Pariat} proposed a null-point and fan-separatrix model driven by the photospheric twisting
motion. As a consequence of the continuous driving by photospheric motions, this model produces recurrent jets,
which explains the quasi-periodicity. \citet{2012Zhang} proposed another possible mechanism that the
null-point reconnection responsible for the quasi-periodicity is modulated by trapped slow-mode waves
along the spine field lines, and slow-mode wave has been reported in, e.g., \citet{2007Cirtain}.
Both models consider magnetic topologies with null-points. But
how to explain those events whose magnetic topologies are QSLs or bald patches? Moreover, the
photospheric motion is believed to be the driver of these recurrent jets. Can we find any indicator
of the quasi-periodicity from the photosphere, such as from the magnetic flux evolution, the velocity field, or the electric current evolution?

In this paper, we present three recurrent coronal jets that were observed on 2010 September 17
by the Atmospheric Imaging Assembly (AIA; \citealt{2012Lemen}) on board the \textit{Solar Dynamics
Observatory} (\textit{SDO}). In order to find a clue on the magnetic reconnection model and the
physical mechanism accounting for the quasi-periodicity, we study the EUV flux, the magnetic flux,
the velocity field, and the vertical electric current evolutions. The vector magnetic fields are
observed by the Helioseismic and Magnetic Imager (HMI; \citealt{2012Scherrer,2012Schou}) on board
\textit{SDO}. Section \ref{sec:data} presents the data analysis and results. Discussion and
conclusion are given in Section~\ref{sec:disc}.

\section{Data Analysis and Results} \label{sec:data}

\subsection{Jets Observed in EUV} \label{sec:jets}

\textit{SDO}/AIA provides high cadence of 12 s, high spatial resolution of $1.5''$ (the pixel
sampling is $0.6''$ per pixel), and high signal-to-noise observations nearly simultaneously
in seven EUV lines, two UV continuums, and one white-light band. Figure~\ref{fig:euv} displays
three composite EUV images showing that three coronal jets recurred on the border of active region (AR) 11106.
All the EUV images have been aligned with ``aia\_prep.pro'' in the Solar SoftWare (SSW).
A line-of-sight magnetic field
observed by \textit{SDO}/HMI is displayed in Figure~\ref{fig:euv}d as an example, which shows
that some parasitic positive polarities distribute in the main negative polarities.
\textit{SDO}/HMI has a spatial resolution of $1''$ (the spatial sampling is $0.5''$ per pixel)
and a cadence of 45~s for the line-of-sight magnetic field. It has also been aligned
with the EUV images with ``aia\_prep.pro''.

\begin{figure}
\centering
\includegraphics[width=0.50\textwidth]{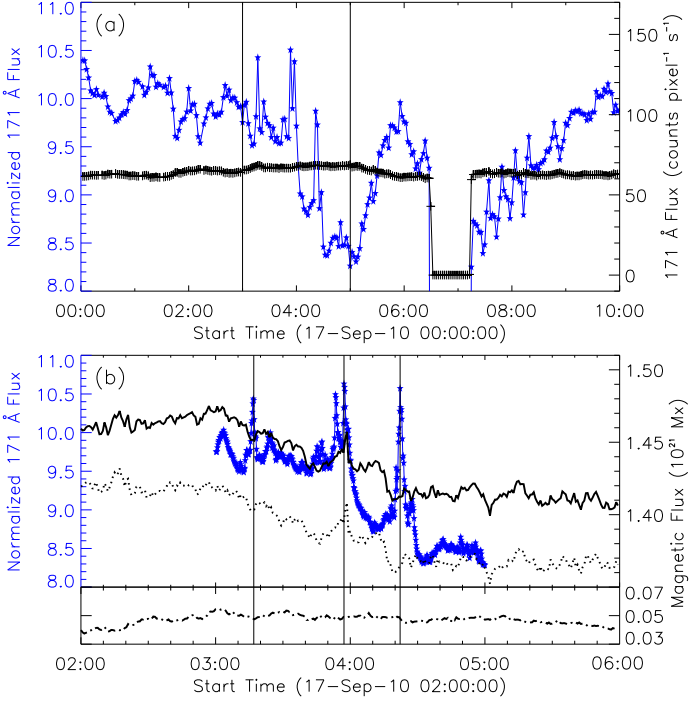}
\caption{\textit{SDO}/AIA 171~\AA \ averaged and normalized flux and \textit{SDO}/HMI line-of-sight magnetic flux.
\textbf{(a)} Blue stars indicate the 171~\AA \ flux in the jet region divided by the number of pixels in the box
(larger solid box as shown in Figure~\ref{fig:euv}a--\ref{fig:euv}c) normalized to the background flux, which is computed
in the dashed box shown in Figure~\ref{fig:euv}. Plus signs represent the averaged background flux per pixel. The temporal resolution
is about 2 minutes. The two vertical lines mark the time range for 171~\AA \ flux as shown in the bottom panel.
\textbf{(b)} The temporal resolution for the 171~\AA \ flux is about 12 s. The three vertical lines mark the peak times of
the 171~\AA \ flux. Dash-dotted, dotted, and solid lines indicate positive, unsigned negative, and total unsigned
line-of-sight magnetic flux, respectively.
} \label{fig:flux}
\end{figure}

We compute the EUV flux in the 171~\AA \ band to give a quantitative representation of the
evolution of the recurrent jets. The 171~\AA \ flux is computed within a rectangle containing
the full jet region as shown by the larger solid box ($130'' \times 80''$) in Figure~\ref{fig:euv}a--\ref{fig:euv}c.
We select two time ranges to compute the 171~\AA \ flux evolution, i.e., a context range
from 00:00 to 10:00 UT and a smaller analysis range centered
on the analyzed jets from 03:00 to 05:00 UT. Due to the limitation of computation resources, the
temporal resolution for the context range is 2 minutes and it is 12 seconds for the analysis range.
The 171~\AA \ flux evolutions for the two time ranges are plotted in Figures~\ref{fig:flux}a and
\ref{fig:flux}b, respectively. The data gap between 06:31 to 07:13 UT is due to the \textit{SDO}
eclipse by the Earth. We have normalized the 171~\AA \ flux to its background level,
which is computed in a quiet region as shown in the dashed box of Figure~\ref{fig:euv}a--\ref{fig:euv}c. The 171~\AA \
background flux shows a flat curve (Figure~\ref{fig:flux}a). Therefore, the variations in the 171~\AA \ flux curve in the jet
region are mainly due to the jets themselves, but not the background.

\begin{figure}
\centering
\includegraphics[width=0.50\textwidth]{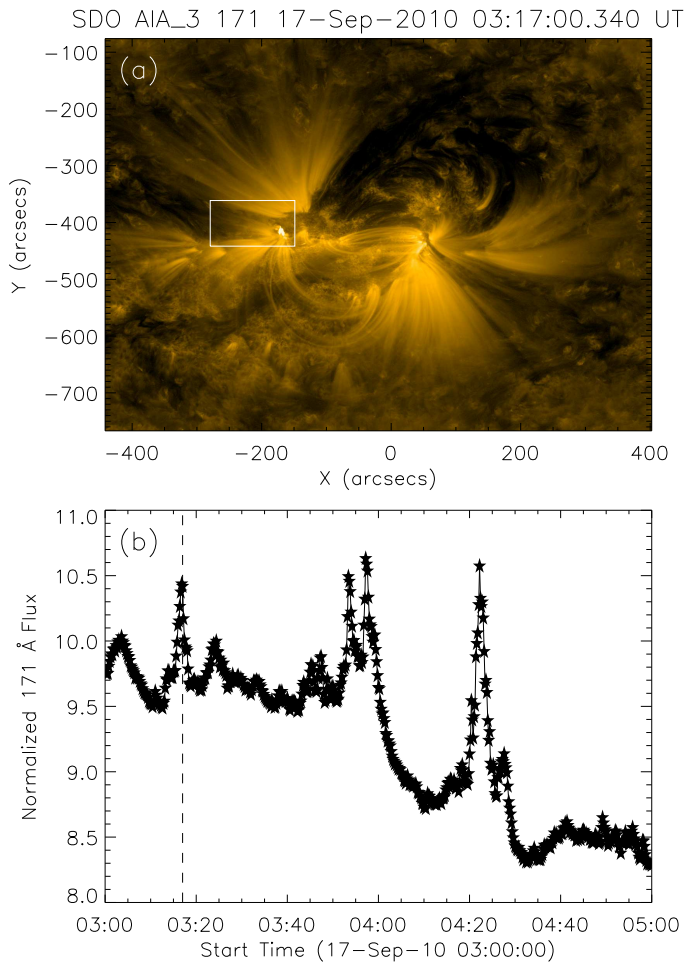}
\caption{Online only figure. A movie showing the evolution of this figure from 03:00 to 05:00 UT
is available in the online journal. \textbf{(a)} \textit{SDO}/AIA 171~\AA \ image. The box marks the region where
the 171~\AA\ fluxes are computed. \textbf{(b)} \textit{SDO}/AIA 171~\AA \ flux normalized to
the background level, which is defined as the flux in the quiet Sun region (see Section~\ref{sec:jets}
for more details). The vertical dashed line marks the time of the 171~\AA \ image in panel (a).
} \label{fig:171}
\end{figure}

Figure~\ref{fig:flux}a shows many fluctuations or peaks in the 171~\AA \ flux curve
during the context range, which indicates that there are many brightening points or jets during this time
range in the selected region. However, the peaks during the analysis range are more distinct than the
others. Therefore, we select this time range for a detailed analysis. The 171~\AA \ flux curve in
Figure~\ref{fig:flux}b has three main peaks at 03:17 UT, 03:57 UT, and
04:22 UT, respectively. There are some smaller peaks before the jet at 03:57 UT.
Using the 171~\AA \ movie (attached to the online-only Figure~\ref{fig:171}), we find
that this jet consists of successive ejections during a short period. The other two events are relatively
 simple with only one ejection in each event.

\subsection{Line-of-Sight Magnetic Flux Evolution} \label{sec:magn}

\begin{figure}
\centering
\includegraphics[width=0.5\textwidth]{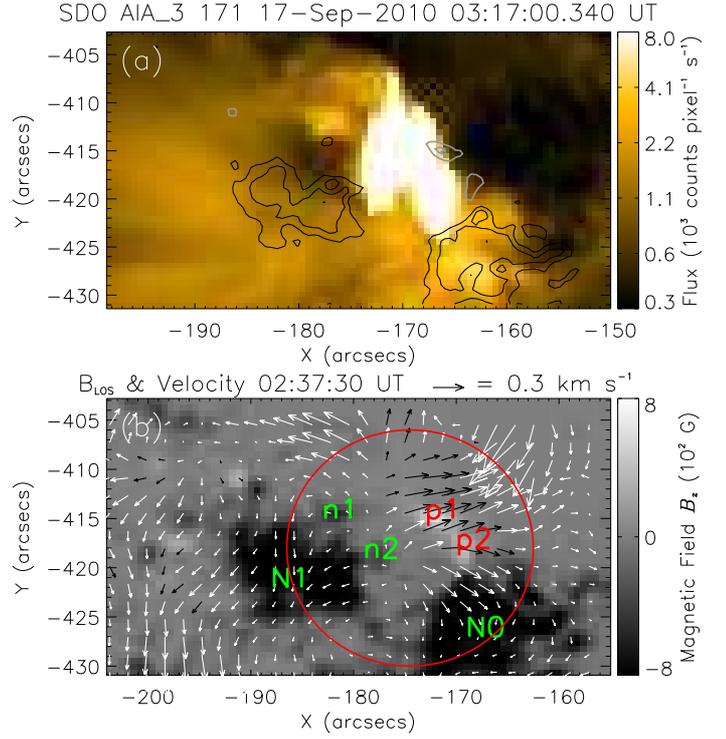}
\caption{\textbf{(a)} \textit{SDO}/AIA 171~\AA \ image overlaid by the line-of-sight magnetic field
observed at the same time as that of the 171~\AA \ image. Gray/black lines mark the contour levels of the
positive/negative polarities. The contour levels of $B_z$ are -1200, -900, -600, 300, and 600 G.
\textbf{(b)} Map of photospheric transverse velocities derived from LCT analysis with
the HMI magnetograms, employing a FWHM correlation window of $5''$ during the time interval 02:25--02:50~UT.
Arrows represent velocities and the background is the line-of-sight magnetogram with positive (white) and
negative (black) polarities. Black/white arrows are located on positive/negative polarities.
North/south is up/down, west/east is towards right/left. The magnetic polarities involved in the jets are
identified with annotations (p1 and p2 for parasitic positive, n1 and n2 for parasitic negative, while N0 and N1
for the main following negative polarities). The circle
highlights the region of interest where the diverging flows are found. A movie showing the evolution of the
line-of-sight magnetic field from 02:00 to 06:00 UT is available in the online journal.
} \label{fig:lct}
\end{figure}

Figure~\ref{fig:flux}b displays the unsigned line-of-sight magnetic flux evolution.
The line-of-sight magnetic flux is integrated in a small rectangle containing mainly the foot-points of the jets
with a size of $48'' \times 28''$ as shown by the solid box in Figure~\ref{fig:euv}d. The box size for the 171~\AA \
flux computation is larger than that for the magnetic flux because, on one hand, jets are events occupying areas in the
corona that are more extended than the triggering magnetic field region in the photosphere below. On the other hand,
the parasitic polarities involved with the jets are surrounded by main polarities of the active region, which
would bias the computation of the line-of-sight magnetic flux if the box is too large. The positive magnetic
flux increases from 02:00 to 03:00 UT and evolves almost steadily from 03:00 to 05:00 UT (Figure~\ref{fig:flux}b,
dash-dotted curve). The negative flux curve is dominated by the main negative polarity of the active region and some
of the flux is close to the border of the selected region (Figure~\ref{fig:euv}d), then, due to convective motions
of the photospheric plasma, magnetic flux might be displaced and no longer captured by the selected area. For this
reason, the unsigned negative flux evolution (Figure~\ref{fig:flux}b, dotted curve) is not necessarily related to
the jetting events. It evolves steadily from 02:00 to 03:00 UT and decreases during the recurrent jets from 03:00 to 05:00 UT.
Comparing the magnetic flux evolution with the 171~\AA \ flux evolution, we do not find a clear relationship between them.
The unsigned magnetic flux in the negative polarity, either decreases (for the jets at 03:17
and 04:22 UT) or increases (for the jet at 03:57 UT) before a jet. There is also no significant evolution
of the positive magnetic flux related to the jets. This result is coherent with a progressive storage
of magnetic energy in the corona, reaching a critical point, and release part of the free magnetic energy
without affecting the magnetic flux crossing the photosphere.

Figure~\ref{fig:lct}a shows that the foot-points of the EUV jets are located where parasitic
polarities appear, as follows. As shown in Figure~\ref{fig:euv}d, AR 11106 consists of a leading positive
and a following negative polarity, which has two strong concentrated polarities N0 and N1. The parasitic
polarities noted as p1, p2, n1, and n2 in Figure~\ref{fig:lct}b are present in the following negative
polarity of AR 11106. The positive parasitic polarities p1 and p2 separate westward from the eastern
negative polarity (n1 and n2). Polarities n1 and n2 have almost merged with the following main polarity
N1 of AR 11106, while p2 is pushed against another part of the following main polarity (N0).
The evolution of these polarities can be seen in the movie of the line-of-sight magnetic field
attached to Figure~\ref{fig:lct} (available in the online journal).

We further study the transverse flows of photospheric magnetic features with local correlation
tracking techniques \citep[][ hereafter LCT]{1988November}. By using a Gaussian tracking window
of full width at half maximum (FWHM) of $5''$, we compute the proper motions of magnetic elements over
the \textit{SDO}/HMI sequence of magnetograms. The window size of $5''$ is appropriate because it is
large enough to contain the coherent structures to correlate, but not too large to keep as much as
possible the spatial resolution and also to limit the computation time to feasible values.
The time series used for this analysis starts at 02:00 UT on 17 September 2010 and is composed by
137 images with a cadence of 45 seconds. Images are grouped in 25-min series for the LCT analysis.
Prior to applying LCT the sequence of images is aligned to eliminate possible jitter and rotation of
the observed target within the field of view. Since the differential rotation is of minor influence
here due to the small field of view, we do not remove this effect. Taking the first image as reference, the
second image is aligned to it by correlating them. Then, the third image is aligned with the second one with
the same method. The process is repeated until all the 137 images are aligned with each other. We find that
the maximum shift is up to 21 pixels in the East-West direction and less than 1 pixel in the North-South
direction. The absolute values of the transverse velocities should be taken with caution because
the LCT technique in general produces some systematic errors in the determination of proper motions
\citep{1988November}. The LCT method may underestimate 20\%--30\% of the velocities in extreme
cases \citep{1994Molowny-Horas}. But it usually generates reliable flow patterns \citep{1988November,2008Vargas}.

The results confirm the emergence pattern seen in the magnetic field movie. There is indeed a
diverging flow with the polarities p1 and p2 moving mostly westward and their negative counterparts,
n1 and n2, moving mostly eastward (this diverging flow pattern is, in average inclined by about $20^\circ$
on the East-West direction), while the bipoles n1-p1 and n2-p2 are mostly East-West oriented at
that time.

\begin{figure*}
\centering
\sidecaption
\includegraphics[width=13cm]{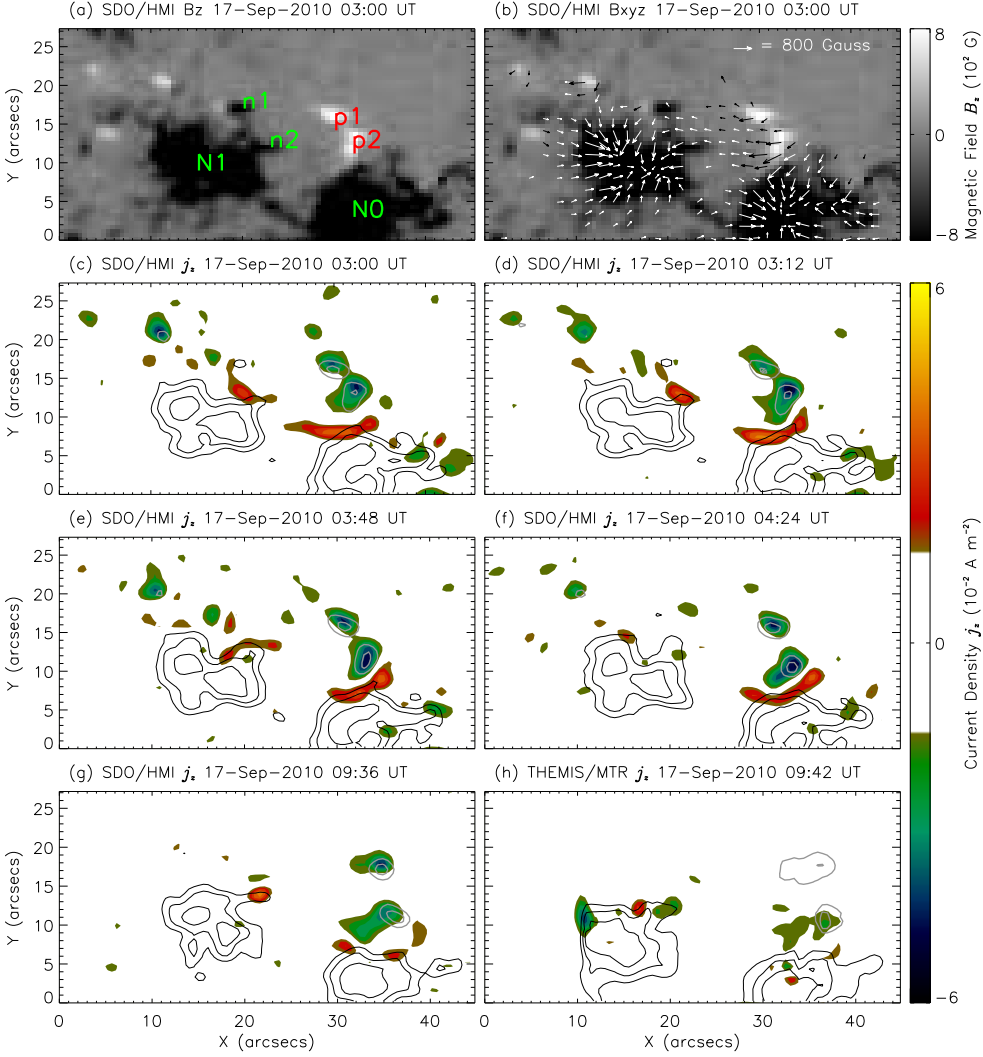}
\caption{\textbf{(a)} \textit{SDO}/HMI vertical component of the magnetic field ($B_{z}$). The
projection effect is removed. \textbf{(b)} \textit{SDO}/HMI vector magnetic field with the
$180^\circ$ ambiguity being removed. \textbf{(c)--(g)} Distribution of the vertical electric
current density $j_z$ averaged over 12 minutes at different times and observed by \textit{SDO}/HMI.
The middle time of the observations is indicated at the top of each panel. Gray/black lines
mark the contour levels of the positive/negative polarities. The contour levels of
$B_{z}$ are $-1200, -900, -600, 300$, and $600$~G. The color bar of $j_z$ is shown on the right side.
\textbf{(h)} Distribution of $j_z$ observed by THEMIS/MTR centered at 09:42 UT. The scan
duration is about 16 minutes in this field of view. The contour levels of $B_{z}$ are $-882, -661,
-440, 110$, and $220$~G, which are at the same percentage of their minimum (for the negative contours)
and maximum (for the positive contours) values as that in Figure~\ref{fig:vector}g.
} \label{fig:vector}
\end{figure*}

\subsection{Vector Magnetic Field} \label{sec:vectormagn}

We further analyze the vector magnetic fields observed by \textit{SDO}/HMI, which records six
filtergrams at each of six wavelengths every 135 s. The filtergrams at each wavelength
correspond six polarization states,
i.e., $I \pm S$, where $S = Q, U$, and $V$, where the notations $I, Q, U,$ and $V$ represent
the four Stokes parameters. In order to increase the signal to noise ratio and remove the
p-mode signals, all the filtergrams are averaged over 12 minutes. More precisely, the average
uses a cosine-apodized boxcar with an FWHM of 720 s. The tapered
temporal window is actually 1215 s. The Stokes parameters, $I$, $Q$,
$U$, and $V$, are computed from the filtergrams at each of the six wavelengths. The vector magnetic
field and other thermodynamical parameters are fitted by the inversion code of Very Fast Inversion
of the Stokes Vector (VFISV; \citealt{2011Borrero}). The $180^\circ$ ambiguity for the transverse
components of the vector magnetic field has been resolved by the improved version of the minimum
energy method \citep{1994Metcalf,2006Metcalf,2009Leka}. We analyze all the 11 vector magnetic fields
between 03:00 to 05:00 UT. The selected field of view for the analysis is $x \in [-300'',190'']$
and $y \in [-560'',-315'']$.

To study the electric current well after the recurrent jets and to compare with other observations,
we analyze two more vector magnetic fields, one observed by \textit{SDO}/HMI at 09:36 UT and the
other one by the T\'elescope H\'eliographique pour l'Etude du Magn\'etisme et des Instabilit\'es
Solaires/Multi-Raies (THEMIS/MTR; \citealt{2000Lopez,2007Bommier}). THEMIS/MTR is a spectro-polarimeter
with higher polarimetry accuracy but lower cadence than that of \textit{SDO}/HMI. THEMIS/MTR
is free from systematic instrumental polarization \citep{2000Lopez}, while the polarimetric accuracy
of \textit{SDO}/HMI is about 0.4\% \citep{2012Schou2}. THEMIS/MTR scanned the solar surface
from east to west to cover the field of view $161'' \times 104''$ in
the time range of 09:34--10:38 UT. The scan step size along east--west direction is about $0.8''$,
the pixel sampling along north--south direction is about $0.2''$. We align the magnetic field observed
by THEMIS/MTR with that by \textit{SDO}/HMI using a common feature identification method. First, a
line-of-sight \textit{SDO}/HMI magnetic field at the middle time (10:06 UT) of the THEMIS/MTR observation
is interpolated to the THEMIS/MTR spatial resolution. Then, we select some common features on both
line-of-sight magnetic fields and record their coordinates. Finally, the offsets of the THEMIS/MTR data
referred to the \textit{SDO}/HMI data are computed with the coordinates recorded in the previous step.

Since the centers of the field of views for the regions of interest are not close to the center of the
solar disk, the projection effect must be removed. We remove this effect with the method
proposed by \citet{1990Gary}, which converts the line-of-sight ($B_\zeta$) and transverse components
($B_\xi$ and $B_\eta$) to the heliographic components ($B_x, B_y$, and $B_z$) and we project those
fields from the image plane to the plane tangent to the solar surface at the center of the field
of view. This coordinate transformation is applied in the field of views given above,
while a smaller region is cut and displayed in Figures~\ref{fig:vector}a and \ref{fig:vector}b as an example.

\subsection{Electric Current Evolution} \label{sec:curr}

Using the vector magnetic field, we compute the vertical electric current density with the Amp{\`e}re's law,
\begin{equation}
j_z(x,y) = \frac{1}{\mu_0}\left ( \frac{\partial B_y}{\partial x} - \frac{\partial B_x}{\partial y} \right ), \label{eqn1}
\end{equation}
where $\mu_0 = 4 \pi \times 10^{-3}$ G m A$^{-1}$. The vertical electric current densities are computed
for all the analyzed vector magnetic fields. Some selected $j_z$ distributions before, close to the three
peaks of, and after the recurrent jets are plotted in Figures~\ref{fig:vector}c--\ref{fig:vector}h.
The $j_z$ distributions close to the jet peaks have very similar patterns as shown
in Figure~\ref{fig:vector}c--\ref{fig:vector}f. However, the magnetic field at 09:36 UT well after the jets has much lower
electric current density (Figure~\ref{fig:vector}g) than those in the jets duration do (Figure~\ref{fig:vector}c--f). The $j_z$
distribution computed by the vector magnetic field observed by THEMIS/MTR as shown in
Figure~\ref{fig:vector}h also shows very low electric currents. The difference between
Figure~\ref{fig:vector}g and \ref{fig:vector}h is caused by the different spatial and temporal resolutions
and spectro-polarimetric accuracies of \textit{SDO}/HMI and THEMIS/MTR. Besides the above reasons,
they are instruments of two different types, where THEMIS/MTR is a spectrograph and \textit{SDO}/HMI is a
filtergraph. THEMIS/MTR records the Stokes line profiles along a slit at one exposure but needs to scan
perpendicularly to the slit to cover a required field of view. While \textit{SDO}/HMI records an image
at a wavelength point at one exposure but needs to scan along the wavelength to get the Stokes line profiles.

\begin{figure}[h]
\centering
\includegraphics[width=0.47\textwidth]{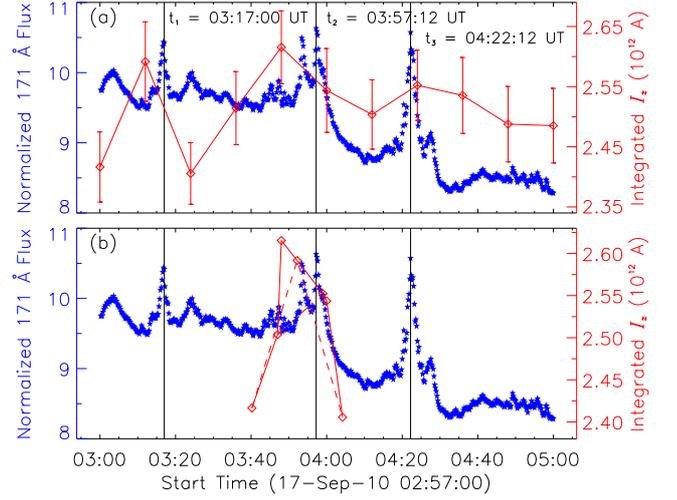}
\caption{\textbf{(a)} Evolutions of the 171~\AA \ flux (normalized as in Figure~\ref{fig:flux}) and the unsigned
vertical electric current $I_z$, which is integrated in the field of view as shown in Figure~\ref{fig:vector}.
Blue stars mark the normalized 171~\AA \ flux and red diamonds mark the unsigned vertical electric current $I_z$.
\textbf{(b)} The electric current evolution curves at the first and third peaks have been shifted to the
second peak and the total profile is shown with a continuous line. The shifted time differences are
$t_2 - t_1$ and $t_3 - t_2$ for the two curves (denoted by dashed and dash-dotted lines), respectively.
Note that the vertical axes for both the 171~\AA \ flux and $I_z$ do not start from zero.
} \label{fig:current}
\end{figure}

Next, we integrate $|j_z|$ in the field of view as shown in Figures~\ref{fig:vector}a--\ref{fig:vector}h
for those regions where $|j_z|$ is above the noise level 0.02~A~m$^{-2}$ estimated by the
formula $(1/\mu_0) (\delta B_\mathrm{T}/\Delta x)$ \citep{1995Gary}, where
$\delta B_\mathrm{T} \sim 100$~G is the assumed transverse field error and $\Delta x = 0.5''$
is the grid size. The field of view in Figures~\ref{fig:vector}a--\ref{fig:vector}h are selected
to be similar to the one used to compute the line-of-sight magnetic flux (Figures~\ref{fig:euv}d and \ref{fig:flux}b).
But they are not identical, since the projection effect has been corrected to compute the vertical
electric currents. Figure~\ref{fig:current}a displays the evolution of the integrated vertical electric current, $I_z$,
where an obvious quasi-periodicity can be found. This recurrent evolution is present on the background of a
stationary current pattern, whose average electric current is $(2.51 \pm 0.06) \times 10^{12}$~A.
The current pattern is stationary since the vertical current density distribution does not change
too much during the jets as shown in Figures~\ref{fig:vector}c--\ref{fig:vector}f.
The fluctuation of $I_z$ around the background (defined as $|\max(I_z) - \min(I_z)|/\bar{I_z}$, where $\bar{I_z}$
denotes the average of $I_z$) is limited within 8\%, which is above the errors of $I_z$ (around 2\%).
The integrated electric currents, $I_z$, for the magnetic fields observed by \textit{SDO}/HMI at 09:36 UT
(Figure~\ref{fig:vector}g) and by THEMIS/MTR at 09:42 UT (Figure~\ref{fig:vector}h), are around
$(1.88 \pm 0.05) \times 10^{12}$~A and $(0.43 \pm 0.04) \times 10^{12}$~A, respectively. Therefore, the currents
significantly decrease later on.

The errors for the integrated $I_z$ as shown in Figure~\ref{fig:current}a are estimated by
a Monte-Carlo method. The basic assumption is that the errors for $I_z$ are normally
distributed in the whole field of view of Figure~\ref{fig:vector} and three times the
standard deviation of the errors is 0.02~A~m$^{-2}$. We add those errors to $I_z$ and integrate
their absolute values with the criteria described in the previous paragraph. After repeating 50
times for each observation time, the errors are estimated as the standard deviation of the 50 integrated $I_z$.

Figure~\ref{fig:current} also shows the integrated vertical electric currents overlaid on the 171~\AA \ flux.
The integrated vertical current is derived from the de-projected heliographic vector maps, while the 171~\AA \
flux is computed in the image plane (plane of the sky). The de-projection allows to derive the
vertical current density, which has a more physical meaning than the line-of-sight one before de-projection. Since the
EUV jets are coronal events and the electric current is on the photosphere, the field of view for computing the
vertical electric current does not need to be the same as that for the 171~\AA \ flux.
We find that the peaks on the $I_z$ curve have a good coincidence with those peaks on the
171~\AA \ flux. They both repeat three times. However, the magnitudes of the peaks on the $I_z$ curve have
large differences, which can be explained by the lower temporal cadence of \textit{SDO}/HMI vector magnetic field
observations (12 minutes). Therefore, sample points on the $I_z$ curve are randomly distributed compared to
the 171~\AA \ flux peak time (Figure~\ref{fig:current}a).

In order to mimic a higher temporal resolution of the $I_z$ evolution, we combine the observations
at the three peaks with one assumption. We assume that the vertical electric current accumulation and
decrease processes are similar for all the three recurrent jets, which are homologous from EUV observations,
i.e., we assume that the peak magnitude of the $I_z$ curve and the evolution timing relative to the 171~\AA \
flux are almost the same. Then, we shift the $I_z$ curve around $t_1 =$ 03:17 and $t_3 =$ 04:22 UT to the
time $t_2 =$ 03:57 UT as shown in Figure~\ref{fig:current}b. From the combined $I_z$ curve in
Figure~\ref{fig:current}b, we find that the vertical electric current
increases first and then decreases in a jet process.

The combined $I_z$ curve in Figure~\ref{fig:current}b also shows two additional features. First,
the $I_z$ curve peak is earlier than the 171~\AA \ flux peak. Second, the vertical electric current
seems to increase faster than it decreases. However, we remind the limited temporal cadence of SDO/HMI,
so both results will need to be checked with other observations.

\begin{figure*}
\centering
\includegraphics[width=1.0\textwidth]{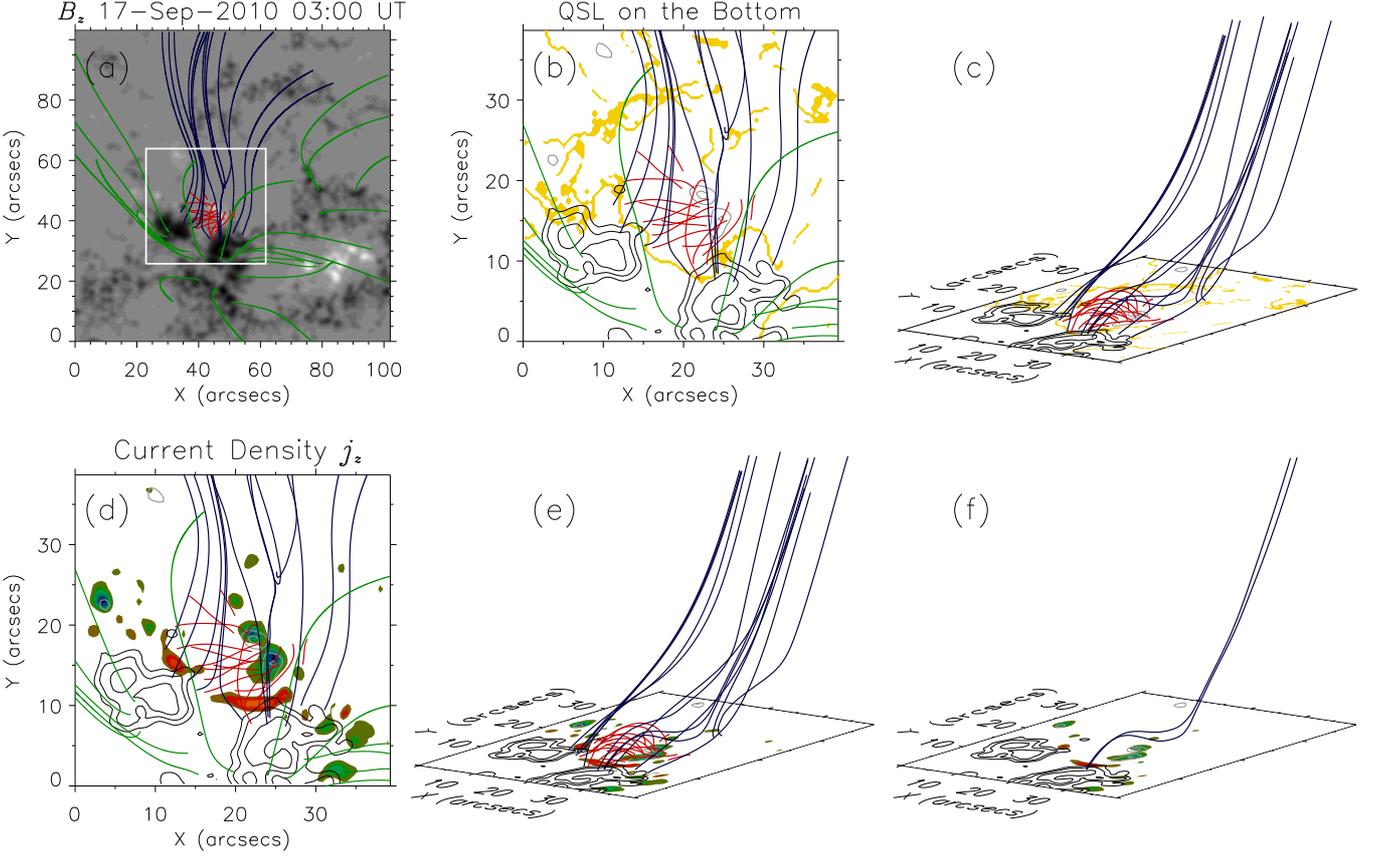}
\caption{\textbf{(a)} Magnetic field lines computed from the nonlinear force-free field model. Different
colors indicate different field line systems. The background is the vertical magnetic field $B_z$
shown with grey levels. White/black color represents positive/negative magnetic polarity. The box
marks the field of view shown in panels (b)--(f). \textbf{(b, c)} Orange ribbons represent the QSL
sections on the photosphere. Only the QSLs with the squashing degree $Q \ge 10^{14}$ are plotted.
Grey/black contours are the same as that in Figure~\ref{fig:vector}c. \textbf{(d, e)} Magnetic
field lines overlaid on the photospheric distribution of the vertical electric current density $j_z$ as
shown in Figure~\ref{fig:vector}c. The current $j_z$ is computed with the non-preprocessed vector magnetic
field. \textbf{(f)} Similar to panels (d) and (e) but only some sample field lines with magnetic dips are shown.
} \label{fig:topo}
\end{figure*}

\subsection{Magnetic Configuration before the Recurrent Jets} \label{sec:nlfff}

The magnetic configuration is one of the necessary information for discriminating different magnetic
reconnection models. For this reason, we study the three-dimensional magnetic field with the nonlinear
force-free field (NLFFF) extrapolation of the coronal field from the vector magnetic field as the boundary
condition. The optimization method is selected for the NLFFF extrapolation \citep{2000Wheatland,
2004Wiegelmann}. The bottom boundary is provided by the \textit{SDO}/HMI vector magnetic field at
03:00 UT, so just before the studied jets. The projection effect has been corrected as described in
Section~\ref{sec:magn}. Since the photospheric vector magnetic field does not satisfy
the force-free condition, we apply the preprocessing method to remove the net magnetic force and torque
on the bottom boundary \citep{2006Wiegelmann}. A requirement for the preprocessing is that, the magnetic field
at the bottom boundary needs to be isolated and flux balanced. Therefore, we select a field of view as shown
by the dotted box in Figure~\ref{fig:euv}d to enclose the major magnetic flux of AR 11106.
Since removing the projection effect changes the geometry of the field of view, we cut off the edges where
the data are incomplete to get a rectangle box, which is resolved by $932 \times 486$ grid points with
$\Delta x = \Delta y \approx 0.5''$. In order to measure the flux balance quantitatively, we define
the flux balance parameter as
\begin{equation}
\epsilon_f = \frac{\sum_i (B_z)_i \mathrm{d} S}{\sum_i |(B_z)_i| \mathrm{d} S} , \label{eqn2}
\end{equation}
where the pixel index $i$ runs all over the field of view, and $\mathrm{d}S$ is the area of one pixel.
The flux balance parameter $\epsilon_f$ is about 0.03 for the magnetic field in this field of view,
which is a tolerably small imbalance.

The boundary field of view for the NLFFF extrapolation is shown in Figure~\ref{fig:topo}a.
There are two reasons why we use this small field of view. First, the region of interest where the jets
originate is small. It is acceptable if we put it in the center. Secondly, the spatial sampling of \textit{SDO}/HMI
is high ($0.5''$). Thus, in the selected field of view (about $100'' \times 100''$) and the height range,
the computation box includes $200 \times 200 \times 200$ grid points, which are reasonable sizes for the
computation. However, \citet{2009Derosa} pointed out that it is critical for the vector magnetic field
to cover as large areas as possible for successful NLFFF modeling. In order to fulfill this requirement
in the limited field of view, we use the following way to prepare the boundary and initial conditions for the higher-resolution
NLFFF extrapolation. First, we compute a lower-resolution NLFFF with the vector magnetic field in the field of view used above
for the preprocessing as shown by the dotted box in Figure~\ref{fig:euv}d. Due to the limitation of computation
resources, the lower-resolution NLFFF is computed in a box of $466 \times 243 \times 201$ grid points with
$\Delta x = \Delta y = \Delta z \approx 1.0''$. Then, we cut out a sub-volume in the field of view as shown
in Figure~\ref{fig:topo}a with a height of $100''$ and interpolate the NLFFF to the spatial resolution of
$\Delta x = \Delta y = \Delta z \approx 0.5''$. The interpolated NLFFF is used as the initial, lateral,
and top boundary conditions for the higher-resolution NLFFF extrapolation.
The bottom boundary is provided by the aforementioned preprocessed vector magnetic field in the
field of view as shown in Figure~\ref{fig:topo}a, which is cut from the larger flux-balanced area.
We use a buffer zone with 25 grid points, in the NLFFF optimization
method, to decrease the effect of these boundaries. The weighting function decreases from 1 to 0 with a cosine
profile in the buffer zone. Finally, we refer to \citet{2000Wheatland} and \citet{2004Wiegelmann} for more
detailed descriptions of the extrapolation method.

We adopt two metrics to test if the NLFFF has reached the force-free and divergence-free state. The
force-free metric is defined as the current-weighted average of $\sin \theta$ \citep{2000Wheatland}:
\begin{equation}
<\mathrm{CW} \sin \theta> = \frac{\sum_i J_i \sin \theta_i}{\sum_i J_i}, \label{eqn3}
\end{equation}
where
\begin{equation}
\sin \theta_i = \frac{| \mathbf{J}_i \times \mathbf{B}_i |}{J_i B_i}, \label{eqn4}
\end{equation}
$\theta$ represents the angle between the current density $\mathbf{J}$ and the magnetic field $\mathbf{B}$,
and $B = |\mathbf{B}|$, $J = |\mathbf{J}|$. The summation is done in the sub-domain of $77 \times 75 \times 35$
grid points, whose projection is shown as the solid box in Figure~\ref{fig:topo}a. Only the grid points where
$J \ge 0.02$~A~m$^{-2}$ (the noise level in Section~\ref{sec:curr}) are considered in computing $\sin \theta$.
For a perfectly force-free magnetic field,
the force-free metric should be zero, i.e., the current density $\mathbf{J}$ and the magnetic field $\mathbf{B}$
are parallel to each other. The divergence-free metric is defined as the unsigned average of the fractional flux
unbalance, $<|f_i|>$, where
\begin{equation}
|f_i| = \frac{|(\nabla \cdot \mathbf{B})_i|}{6 B_i / \Delta x}. \label{eqn5}
\end{equation}
The divergence-free metric is computed in the same domain as that for the force-free metric, while the difference
is that all the grid points of the sub-domain are considered. This metric should be zero in a truly divergence-free field.

The force-free and divergence-free metrics for the NLFFF model are $0.19$ and $3.0 \times 10^{-3}$,
respectively. The corresponding angle for the force-free metric, which is defined as $\arcsin <\mathrm{CW} \sin \theta>$,
is $11.0^\circ$. \citet{2008Metcalf} compared various NLFFF algorithms using simulated
chromospheric and photospheric vector fields. The force-free metric $0.19$ that we derive here is slightly better
than that of 0.26 using the preprocessed and smoothed photospheric boundary with the optimization code of
\citet{2004Wiegelmann}. However, it is larger, therefore worse, than the force-free metric 0.11 using the
chromospheric boundary. The divergence-free metric $3.0 \times 10^{-3}$ is larger (and worse) than both the
metrics using the chromospheric and preprocessed and smoothed photospheric boundaries with the optimization
method as listed in Table 5 of \citet{2008Metcalf}, since they adopted simulated vector magnetic fields
as boundary conditions, which are smoother than the observational boundaries that we use in this paper.

Some selected magnetic field lines in the NLFFF model are plotted in Figure~\ref{fig:topo}. The overall
magnetic configuration in Figure~\ref{fig:topo}a shows different connectivities with small closed field lines
linking the small diverging bipole (represented by the red lines connecting positive polarities p1 and p2 to
negative ones n1 and n2) and long field lines (represented by the blue lines anchored in polarities N0 and N1).
We find no magnetic null point in between these two set of field lines, so no separatrix. Rather, a continuous
but drastic change of magnetic connectivity, so a QSL, is separating both types of connectivities. We compute
the squashing degree $Q$ as defined in \citet{2002Titov} to locate the position of the QSLs. Figures~\ref{fig:topo}b
and \ref{fig:topo}c show the QSL sections on the bottom boundary. The closed (red) and long (blue) field lines
are clearly separated by QSLs between them.

We also overlay the magnetic field lines on the photospheric distribution of the vertical electric current
density $j_z$ as shown in Figures~\ref{fig:topo}d and \ref{fig:topo}e. The footpoints of those field lines
close to the QSL are rooted on the regions where $j_z$ is large. This result is coherent with previous findings
where concentrated electric currents have been found at the border of the QSLs in flare configurations
\citep[see][ and references therein]{1997Demoulin}. For a thin QSL, a moderate difference of magnetic stress across
the QSL creates strong electric current. Then, a current layer is easily build up at a QSL. For the first time,
the high temporal cadence and magnetic sensitivity of \textit{SDO}/HMI allows to diagnose such current layer
evolution for such small events as the studied jets.

Due to the existence of the parasitic positive magnetic polarities, some blue field lines are bent down
from higher altitudes to lower altitudes. Magnetic dips and bald patches are present where the magnetic field
lines are bent down as shown in Figure~\ref{fig:topo}f. We refer to the following theoretical or
observational papers for detailed definition and analysis of magnetic dips and bald patches, such as
\citet{1993Titov}, \citet{1996Bungey}, \citet{2002Mandrini}, and \citet{2005Demoulin}. However, it is not
clear if the bald patches played any role in the magnetic reconnection of the studied jets, which asks for further studies.

\section{Discussion and Conclusion} \label{sec:disc}

In summary, we find that three EUV jets observed by \textit{SDO}/AIA from 03:00 to 05:00 UT
on 2010 September 17 recurred in the same region close to the border of AR 11106.
The time differences between the three successive 171~\AA \ peaks are 40 minutes and 25 minutes.
On the border of the following negative polarity, parasitic positive polarities (polarities p1 and p2
in Figure~\ref{fig:lct}) are present close to the footpoints of the EUV jets. They are parts of a
magnetic bipole which is growing in size as shown by the diverging flow pattern. The line-of-sight
magnetic flux evolution does not have a clear relationship with the 171~\AA \ flux evolution. However,
the high time cadence of \textit{SDO}/HMI (12 minutes for the vector magnetic field) allows us to follow the photospheric current
evolution. We integrate the absolute vertical electric current density found in the region of the jets.
The peaks of this total current have a good temporal coincidence with the peaks of the 171~\AA\ flux.

The aforementioned features can only be partly explained by the emerging flux model \citep{1992Shibata265}
and the converging flux model \citep{1994Priest} since the newly emerged magnetic flux is consistent
with the former model and the bald patch configuration is consistent with the latter one. But both models are
two dimensional with magnetic reconnection in separatrices (as implied by their dimensionality), while
the magnetic connectivity is not necessarily discontinuous in the three dimensional space. The above models
can be generalized to three dimensional configuration having a magnetic null point
\citep[e.g.,][]{2008Moreno-Insertis,2009Torok,2009Pariat,2010Pariat}.
However, in the present study, no null point is found in the corona with the NLFFF extrapolation. Rather
continuous changes of connectivity are present with drastic changes at some locations. Indeed, the
computed coronal configuration has a QSL separating a small-scale evolving double bipole (polarities p1--n1
and p2--n2 in Figure~\ref{fig:lct}) from the large scale AR magnetic field, a configuration comparable to
those found previously, e.g. in an X-ray bright point \citep{1996Mandrini}.

The temporal evolution of the photospheric magnetic field (see the movie attached to Figure~\ref{fig:lct}) as well as the velocities
deduced by LCT both indicate divergent flows in the small parasitic bipoles. These flows would force the closed field
lines of the small parasitic bipole to grow in size, and to interact with the overlaying large scale field lines
(Figure~\ref{fig:topo}). Such kind of evolving magnetic configuration is known to build up a narrow current
layer in the QSL as conjectured analytically \citep{1996Demoulin} and found in numerical simulations even
for relatively small displacements \citep{2005Aulanier,2011Effenberger}. The main reason is
that the magnetic stress of very distant regions, generated by photospheric flows, are brought close to one
another, typically over the QSL thickness. This current layer becomes thinner and stronger with time. When the
currents reach the dissipative scale, there is a breakdown of ideal MHD in the QSLs, magnetic reconnection
occurs and part of the magnetic energy is released \citep{2006Aulanier}. In the above studied configuration,
this process would create some newly formed lower-lying field lines with less shear and some newly formed higher
field lines, along which hot collimated plasma are ejected into the higher corona.

The quasi-periodicity and homology of the coronal jets requires two necessary conditions. First, the magnetic
energy is injected into the corona uninterruptedly in almost the same region, and secondly, it releases
intermittently. The first condition is satisfied by the observed diverging photospheric motions which inject
free magnetic energy in the coronal field without changing too much the photospheric magnetic field
distribution. The second condition requires the current layer to become thin enough recurrently in order
to reconnect and release rapidly part of the plasma trapped in the low lying loops (of the diverging bipole).
The observed periodic build up and decrease of the photospheric currents, in phase with the EUV jets
(Figure~\ref{fig:current}), is an evidence that this second condition is met. Indeed, only a small fraction
of the current is dissipated, an indication that the reconnection stops rapidly when the current layer weakens.
Then, the next build up of current starts from a non-potential configuration. This allows a relatively fast
build up of a new thin current layer and may explain the small time interval between the jets (40 and 25
minutes). The photospheric flows displace the magnetic polarities by only about 300--480~km during the successive jets,
where we use $v=0.2$~km s$^{-1}$ estimated by the LCT velocity data as the average velocity in the encircled
region of Figure~\ref{fig:lct}.

What is still surprising is that the build-up and decay times of the currents are comparable (Figure~\ref{fig:current}).
Usually, the build-up time is thought to be much longer than the decay time for many solar activities,
such as flares and coronal mass ejections (CMEs). A reason for the comparable time scale could be the limited
time cadence of \textit{SDO}/HMI including the constrain of a large enough signal to
noise ratio, so this kind of study needs a higher cadence and better polarimetric accuracy to get a more
precise temporal evolution of the currents. We mimic a higher temporal cadence by supposing that the evolution
processes are the same, then we combine the current measurements of
the three jets using the EUV flux maximum as a time reference. With these limits, we find a current maximum near
the beginning of the EUV flux rise, about ten minutes before the EUV flux maximum. This indicates that the free
magnetic energy starts to decrease as soon as there is an evidence of reconnection.

In conclusion, we find evidence indicating that the studied recurrent coronal jets are caused by
magnetic reconnection in the QSL present between a diverging bipole and the main AR magnetic field.
Magnetic energy is injected by the continuous diverging motions observed at the photospheric level.
We assume that this process has build a thin current layer whose
photospheric cross sections is deduced from the vector magnetograms. The current evolution is found in phase
with the EUV flux with a current maximum present before the jets followed by a relaxation to lower value.
However, this evolution of the current intensity is limited to around 8\% of the average background.
Note that only the vertical component of electric currents are derived on the photosphere. If the
NLFFF assumption is valid, the coronal currents are directly related to the photospheric currents, since
$\alpha = \mu_0 j_z / B_z$ and $\alpha$ is constant along a field line. Therefore, we conclude that
the magnetic system always stays close to the resistive instability of the current layer.
The jets does not change the photospheric vertical current distribution too much and only a moderate photospheric
evolution is needed to restart a new jet.

All together, we interpret these observations as the consequence of the continuous photospheric driving
of the coronal field. A thin current layer is built in the QSL, where the magnetic field reconnects when
the electric currents are strong enough and stops
reconnecting soon later on. The magnetic reconnection dissipates only a small part of the electric currents.
The above phenomena repeat in time and drive the recurrent jets as the confined plasma in the closed loops
is accelerated, after reconnection, into the large scale field.

\begin{acknowledgements}
We thank the anonymous referee very much for his/her constructive comments that improve this paper.
Data are courtesy of \textit{SDO} and the HMI and AIA science teams. THEMIS is a French telescope
operated by CNRS on the island of Tenerife in the Spanish Observatorio del Teide of the Instituto
de Astrof\'isica de Canarias. We thank the THEMIS team for the observations.
YG and MDD were supported by the National Natural Science Foundation of China (NSFC) under the grant
numbers 11203014, 10933003, 10878002, and the grant from the 973 project 2011CB811402.
BS thanks the team of the flux emergence workshop lead by K. Galsgaard and F. Zuccarello for fruitful discussions in Bern at ISSI.
\end{acknowledgements}


\end{document}